\documentstyle[12pt]{article}
\def \Bbar{\overline{B}}
\def \Kbar{\overline{K}}
\def \s{\sqrt{2}}
\def \st{\sqrt{3}}
\def \sx{\sqrt{6}}
\begin{document}
\rightline{TECHNION-PH-94-8}
\rightline{UdeM-LPN-TH-94-193}
\rightline{EFI-94-12}
\rightline{hep-ph/9404283}
\rightline{April 1994}
\bigskip
\bigskip
\centerline{\bf DECAYS OF $B$ MESONS}
\centerline{\bf TO TWO LIGHT PSEUDOSCALARS}
\bigskip
\centerline{\it Michael Gronau}
\centerline{\it Department of Physics}
\centerline{\it Technion -- Israel Institute of Technology, Haifa 32000,
Israel}
\medskip
\centerline{and}
\medskip
\centerline{\it Oscar F. Hern\'andez\footnote{e-mail:
oscarh@lps.umontreal.ca} {\rm and} David London\footnote{e-mail:
london@lps.umontreal.ca}}
\centerline{\it Laboratoire de Physique Nucl\'eaire}
\centerline{\it Universit\'e de Montr\'eal, Montr\'eal, PQ, Canada H3C 3J7}
\medskip
\centerline{and}
\medskip
\centerline{\it Jonathan L. Rosner}
\centerline{\it Enrico Fermi Institute and Department of Physics}
\centerline{\it University of Chicago, Chicago, IL 60637}
\bigskip
\bigskip
\centerline{\bf ABSTRACT}
\medskip
\begin{quote}
The decays $B \to PP$, where $P$ denotes a pseudoscalar meson, are
analyzed, with emphasis on charmless final states. Numerous triangle
relations for amplitudes hold within SU(3) symmetry, relating (for example)
the decays $B^+ \to \pi^+ \pi^0,~ \pi^0 K^+$, and $\pi^+ K^0$. Such
relations can improve the possibilities for early detection of
$CP$-violating asymmetries. Within the context of a graphical analysis of
decays, relations are analyzed among SU(3) amplitudes which hold if some
graphs are neglected.  One application is that measurements of the rates
for the above three $B^+$ decays and their charge-conjugates can be used to
determine a weak CKM phase. With measurements of the remaining rates for
$B$ decays to $\pi \pi,~\pi K$, and $K \bar K$, one can obtain two CKM
phases and several differences of strong phase shifts.
\end{quote}
\newpage
\centerline{\bf I. INTRODUCTION}
\bigskip

The present explanation of $CP$ violation in the neutral kaon system makes
use of phases in the Cabibbo-Kobayashi-Maskawa (CKM) matrix describing the
charge-changing weak transitions of quarks. The $CP$-violating decays of
mesons (``$B$'' mesons) containing a $b$ or $\bar b$ quark provide numerous
ways to check this picture.

One class of $CP$-violating asymmetries involves the decays of neutral $B$
mesons to $CP$ eigenstates such as $J/\psi K_S$ or $\pi^+ \pi^-$. The
observation of such asymmetries requires one to know whether the neutral
$B$ meson was a $B^0~(\equiv \bar b d)$ or $\bar B^0~(\equiv b \bar d)$ at
the time it was produced. The technical difficulties of ``tagging'' the
flavor of the initial $B$ meson have given rise to a number of suggestions,
including production in energy-asymmetric electron-positron colliders
(``$B$ factories'') or the use of correlations with pions produced nearby
in phase space. Differences in the rates for $B^0|_{t=0}$ and $\bar
B^0|_{t=0}$ decaying to a $CP$ eigenstate arise as a result of the
interference between a direct decay amplitude and one which proceeds
through $B^0 - \bar B^0$ mixing. These rate differences are particularly
easy to interpret if a single weak subprocess contributes to each direct
decay amplitude. In that case, the $CP$-violating asymmetries provide
direct information on phases of CKM matrix elements.

In decays to $CP$ eigenstates, the possibility of two distinct weak
subprocesses contributing to each direct decay amplitude requires a
somewhat more elaborate analysis. As an example, in the case of $B^0 \to
\pi^+ \pi^-$, the dominant subprocess is expected to be one involving $\bar
b \to \bar d u \bar u$. However, if direct $\bar b \to \bar d$
(``penguin'') transitions mediated by virtual $u,~c$, and $t$ quarks and
gluons also contribute, the observed $CP$-violating asymmetries are
affected [1]. Here, the study of rates for all possible charge states in $B
\to \pi \pi$ and $\bar B \to \pi \pi$ decays and the time-dependence of
$B^0(t) \to \pi^+ \pi^-$ allows one to separate by isospin all the relevant
effects and to obtain information on CKM phases [2].

Another class of $B$ decays which are potentially interesting for $CP$
studies includes ``self-tagging'' modes such as $B^+ \to \pi^0 K^+$ and
$B^+ \to \pi^+ K^0$, in which a difference between the branching ratio for
the mode and its charge conjugate immediately signals $CP$ violation. In
order that a difference in rates be observable, it is necessary to have a
non-zero difference between the phase shifts in two strong eigenchannels
($I = 1/2$ and $I = 3/2$ in the case of $\pi K$), as well as two different
weak subprocesses contributing to the decay. In $B \to \pi K$ the two weak
subprocesses are $\bar b \to \bar s u \bar u$ and the CKM-favored $\bar b
\to \bar s$ penguin amplitude. Using isospin, it is possible to separate
these two subprocesses in order to extract the CKM phase from decays of
neutral $B$ mesons to the $CP$ eigenstate $\pi^0 K_S$. For this purpose one
must measure the time-dependence of $B^0(t) \to \pi^0 K_S$, as well as the
rates for decays of all possible charge states in $B \to \pi K$ and $\bar B
\to \pi \bar K$ processes [3-5].

Information about final state phases {\it per se} has already been obtained
in decays of charmed mesons to $\pi \bar K$. By comparing the rates for
$D^0 \to \pi^+ K^-$, $D^0 \to \pi^0 \bar K^0$, and $D^+ \to \pi^+ \bar
K^0$, for example, one can conclude that there is a significant phase
difference between the $I = 1/2$ and $I = 3/2$ elastic $\pi \bar K$
scattering amplitudes at $E_{\rm c.m.} = M_D$ [6]. The same is true for the
decays $D \to \pi \bar K^*$, while no significant phase difference has been
found between the $I = 1/2$ and $I = 3/2$ amplitudes in $D \to \rho \bar K$
[6].

The simplest system in which early information about final state phases in
$B$ decays may be forthcoming is the full set of charge states for $B \to
\pi \bar D$. Rates for $B^+ \to \pi^+ \bar D^0$ and $B^0 \to \pi^+ D^-$
have already been measured, while only an upper bound exists for $B^0 \to
\pi^0 \bar D^0$ [7]. These rates determine the three sides of a triangle;
if the triangle has non-zero area, there exists a phase difference between
the $I = 1/2$ and $I = 3/2$ amplitudes, just as in $D \to \pi \bar K$. In
contrast to the case of $B^0 \to \pi^+ \pi^-$ and $\bar B^0 \to
\pi^+\pi^-$, such a phase difference would not lead to a difference in
rates between such decays as $B^0 \to \pi^+ \bar D^0$ and $\bar B^0 \to
\pi^- D^0$, since a single weak phase contributes in these processes.

Decays such as $B \to \pi \pi$ and $B \to \pi K$ are related to one another
by SU(3) symmetry [8-13]. It is then natural to ask whether the isospin
analyses of Refs.~[2-5] can be generalized to SU(3) in order to obtain
further information about the prospects for observing $CP$ violation in
such decays as $B \to \pi K$, or for measuring weak and final-state phases.
In the present paper we report the results of efforts to find such
information.

We have undertaken a systematic review of the SU(3) predictions for the
decays $B \to P P$, where $P$ stands for a pseudoscalar meson, emphasizing
final states with zero charm. We have examined a number of linear relations
among amplitudes which follow from previous SU(3) analyses, but whose
simplicity seems to have gone unnoticed. Among these is a relation between
the amplitude for $B^+ \to \pi^+ \pi^0$ and pairs of amplitudes for $B \to
\pi K$. This relation is expressed as a triangle in the complex plane. It
is a necessary (but not sufficient) condition for the observability of $CP$
violation in $B^\pm \to \pi K$ decays that this triangle have nonzero area.
The triangle relation simplifies the analysis used to obtain information on
CKM angles from the study of time-dependent neutral $B$ decays to the $CP$
eigenstate $\pi^0 K_S$, which uses a somewhat more involved quadrangle
relation [3-5].

This triangle is also of great utility in another respect. Using only
charged $B$ decays, it allows the clean extraction of a weak CKM phase.
Furthermore, using it and other triangle relations involving the decays $B
\to \pi\pi$ and $B\to \pi K$, along with the rates for $B\to K \bar K$, one
can also obtain information about two weak CKM phases, as well as about
final-state phases. A possible experimental advantage of these methods for
obtaining CKM phase information is that no time-dependent measurements are
necessary -- only decay rates are needed. We have also found relations
between certain final-state phase differences in $B \to \pi\pi$ and $B \to
\pi K$ decays.

Our analysis is performed using a simple graphical method which has been
shown equivalent [8] to a decomposition in terms of SU(3) reduced matrix
elements. The graphical description is overcomplete in the sense that we
can write six different graphs, but they always appear in the form of five
linear combinations, corresponding to the five reduced matrix elements in
an SU(3) decomposition.

It is possible that the graphs which we use to construct decay amplitudes
take on a more direct meaning. Thus, a process which could take place only
as a result of a certain graph (such as quark-antiquark annihilation) might
not be fed by rescattering from another graph. We thus make a systematic
study of the effects of neglecting certain graphs whose contributions are
expected to be small in the limit of infinite mass of the $b$ quark. These
correspond to $W$ exchange, $W$ in the direct channel (``annihilation''),
and annihilation through a penguin graph (vacuum flavor quantum numbers in
the direct channel). We are left with three other types of graph and,
correspondingly, three independent combinations of reduced matrix elements.

It is thus our purpose to draw attention to a number of interesting
questions about {\it amplitudes} and their phases which can be addressed
purely from the standpoint of {\it rates} in $B \to PP$ decays. A full
discussion of amplitudes was performed in Ref.~[8], but without much
emphasis on the simple linear relations among them. The treatment in
Ref.~[9] dealt primarily with rates, for which few direct relations exist
in the $\Delta C = 0$ sector. There does not appear to have been a previous
discussion of the effects of neglect of certain contributions in the
graphical description of decays.

In Section II we recapitulate the full SU(3) analysis of Ref.~[8] in terms
of graphical contributions. We stress the wide variety of relations that
hold among different amplitudes. These include not only separate relations
among $\Delta S = 0$ and $|\Delta S| = 1$ transitions, but also relations
{\it between} the two sectors. We give specific examples relating to $B \to
\pi \pi$ and $B \to \pi K$ decays, and show how to extract useful
final-state interaction information from these processes. For completeness,
we also quote results involving an octet $\eta$, which we denote $\eta_8$,
and mention their limited usefulness. While we concentrate on $\Delta C =
0$ transitions, we briefly treat decays to final states involving charm, in
the context of extraction of final state phase differences from decay
rates.

We next specialize, in Section III, to the case in which certain diagrams
contributing to decay amplitudes are neglected. We translate this
assumption into linear relations among SU(3) reduced matrix elements, and
discuss the corresponding relations among amplitudes for decays. The
physical consequences for observability of $CP$ violation in various
channels are mentioned. The neglect of some diagrams permits one to
determine a weak CKM phase just from the decays of charged $B$ mesons to
$\pi \pi$ and $\pi K$. With measurements of the remaining rates for $B$
decays to $\pi \pi,~\pi K$, and $K \bar K$, one can obtain two CKM phases
and several differences of strong phase shifts.

With appropriate warnings, we treat the case of a ``physical'' $\eta$ and
$\eta'$ using graphical methods in Sec.~IV. Some possible effects of SU(3)
breaking are mentioned in Sec.~V. We summarize our results and discuss
experimental prospects in Sec.~VI. An Appendix lists the decomposition of
reduced matrix elements labelled by SU(3) representations into graphical
contributions, in order to display explicitly the connection between the
two languages.
\bigskip

\centerline{\bf II. SU(3) ANALYSIS}
\bigskip

\leftline{\bf A.  Definitions and counting of reduced matrix elements}
\bigskip

Adopting the same conventions as Ref.~[8], we take the $u,~d$, and $s$
quark to transform as a triplet of flavor SU(3), and the $-\bar u,~\bar d$,
and $\bar s$ to transform as an antitriplet. The mesons are defined in such
a way as to form isospin multiplets without extra signs. Thus, the pions
will belong to an isotriplet if we take
\begin{equation}
\pi^+ \equiv u \bar d~~,~~~\pi^0 \equiv (d \bar d - u \bar u)/\s~~,~~~
\pi^- \equiv - d \bar u~~~,
\end{equation}
while the kaons and antikaons will belong to isodoublets if we take
\begin{equation}
K^+ \equiv u \bar s~~,~~~K^0 \equiv d \bar s~~~,
\end{equation}
\begin{equation}
\bar K^0 \equiv s \bar d~~,~~~K^- \equiv - s \bar u~~~.
\end{equation}
We choose $\eta_8 \equiv (2 s \bar s - u \bar u - d \bar d)/\sx$. The true
$\eta$ looks more like an octet-singlet mixture with a mixing angle of
around 19 to 20 degrees [14], such that $\eta \approx (s \bar s - u \bar u
- d \bar d)/\st$. In order to treat such a state correctly, we would have
to introduce additional reduced matrix elements in the context of SU(3),
additional graphs within the context of a graphical analysis, and decays
involving $\eta'$. We expect the predictive power of such an approach would
be minimal. Thus, we quote results for octet $\eta$'s mainly for
completeness.

The $B$ mesons are taken to be $B^+ \equiv \bar b u$, $B^0 \equiv \bar b
d$, and $B_s \equiv \bar b s$. Their charge-conjugates are defined as $B^-
\equiv - b \bar u$, $\bar B^0 \equiv b \bar d$, and $\bar B_s \equiv b \bar
s$.

We now count reduced matrix elements for transitions to charmless $PP$
final states. The weak Hamiltonian operators associated with the
transitions $\bar b \to \bar q u \bar u$ and $\bar b \to \bar q$ ($q = d$
or $s$) can transform as a ${\bf 3^*}$, ${\bf 6}$, or ${\bf 15^*}$ of
SU(3). When combined with the triplet light quark in the $B$ meson, these
operators then lead to the following representations in the direct channel:
\begin{equation}
{\bf 3^*} \times {\bf 3} = {\bf 1} + {\bf 8}_1~~~,
\end{equation}
\begin{equation}
{\bf 6} \times {\bf 3} = {\bf 8}_2 + {\bf 10}~~~,
\end{equation}
\begin{equation}
{\bf 15^*} \times {\bf 3} = {\bf 8}_3 + {\bf 10^*} +{\bf 27}~~~.
\end{equation}

We are concerned with couplings of these representations to the symmetric
product of two octets (the pseudoscalar mesons, which are in an S-wave
final state). Since ${\bf (8 \times 8)_s} = {\bf 1 + 8 + 27}$, the singlet,
octet, and ${\bf 27}$-plet each have unique couplings to this pair of
mesons, while the decimets cannot couple to them. Thus, the decays are
characterized by one singlet, three octet, and one ${\bf 27}$-plet
amplitude. The decomposition in terms of these amplitudes is given in
Ref.~[8], and is implied by the results of the Appendix. Separate
amplitudes apply to the cases of strangeness-preserving and
strangeness-changing transitions. As we shall see, there are relations
between linear combinations of the strangeness-preserving and
strangeness-changing amplitudes.
\bigskip

\leftline{\bf B. Amplitudes in terms of graphical contributions}
\bigskip

The SU(3) analysis of $\Delta C = 0$ $B \to PP$ decays is equivalent to a
decomposition of amplitudes in terms of graphical contributions. We shall
adopt a notation in which an unprimed amplitude stands for a
strangeness-preserving decay, while a primed contribution stands for a
strangeness-changing decay. The relevant graphs are illustrated in Fig.~1.
They consist of the following [10]:

\begin{itemize}

\item[{1.}] A (color-favored) ``tree'' amplitude $T$ or $T'$, associated
with the transition $\bar b \to \bar q u \bar u$ ($q = d$ or $s$) in which
the $\bar q u$ system forms a color-singlet pseudoscalar meson while the
$\bar u$ combines with the spectator quark to form the other pseudoscalar
meson;

\item[{2.}] A ``color-suppressed'' amplitude $C$ or $C'$, associated with
the transition $\bar b \to \bar u u \bar q$ in which the $\bar u u$ system
is incorporated into a neutral pseudoscalar meson while the $\bar q$
combines with the spectator quark to form the other meson;

\item[{3.}] A ``penguin'' amplitude $P$ or $P'$ associated with the
transition $\bar b \to \bar q$ involving virtual quarks of charge 2/3
coupling to one or more gluons in a loop;

\item[{4.}] An ``exchange'' amplitude $E$ or $E'$ in which the $\bar b$
quark and an initial $q$ quark in the decaying (neutral) $B$ meson exchange
a $W$ and become a $\bar u u$ pair;

\item[{5.}] An ``annihilation'' amplitude $A$ or $A'$ contributing only to
charged $B$ decay through the subprocess $\bar b u \to \bar q u$ by means
of a $W$ in the direct channel; and

\item[{6.}] A ``penguin annihilation'' amplitude $PA$ or $PA'$ in which an
initial $\bar b q$ state annihilates into vacuum quantum numbers.

\end{itemize}

This set of amplitudes is over-complete. The physical processes of interest
involve only five distinct linear combinations of the amplitudes, which are
given in terms of SU(3) direct-channel representations in the Appendix.
Here we shall simply mention acceptable linearly independent sets of
amplitudes once we have expanded all processes of interest in terms of the
above contributions.

The results for $\Delta S = 0$ transitions are shown in Table 1, while
those for $|\Delta S| = 1$ transitions are shown in Table 2. We can
identify the following linearly independent combinations of $\Delta S = 0$
amplitudes, for example:

\renewcommand{\arraystretch}{1.2}
\begin{table}
\caption{Decomposition of $B \to PP$ amplitudes for $\Delta C = \Delta S =
0$ transitions in terms of graphical contributions.}
\begin{center}
\begin{tabular}{l l c c c c c c} \hline
         & Final &  $T$  &  $C$  &  $P$  &  $E$  &  $A$  &  $PA$  \\
         & state &       &       &       &       &       &        \\ \hline
$B^+\to$ & $\pi^+\pi^0$ & $-1/\s$ & $-1/\s$ & 0 &  0   &  0 & 0 \\
         & $K^+ \bar K^0$ & 0 & 0 & 1    &   0   &   1   &   0    \\
         & $\pi^+\eta_8$ & $-1/\sx$ & $-1/\sx$ & $-2/\sx$ & 0 & $-2/\sx$ &
0 \\ \hline
$B^0\to$ & $\pi^+\pi^-$ & $-1$ & 0 & $-1$ & $-1$ &   0   &  $-1$  \\
         & $\pi^0\pi^0$ & 0 & $-1/\s$ & $1/\s$ & $1/\s$ & 0 & $1/\s$ \\
         & $K^+K^-$ & 0  &   0   &   0   &  $-1$ &   0   &  $-1$  \\
         & $K^0 \bar K^0$ & 0 & 0 &  1   &   0   &   0   &   1    \\
         & $\pi^0\eta_8$ & 0 & 0 & $-1/\st$ & $1/\st$ & 0 &  0    \\
         & $\eta_8\eta_8$ & 0 & $1/3\s$ & $1/3\s$ & $1/3\s$ & 0 & $1/\s$
\\ \hline
$B_s\to$ & $\pi^+K^-$ & $-1$ & 0 &  $-1$ &   0   &   0   &   0    \\
         & $\pi^0 \bar K^0$ & 0 & $-1/\s$ & $1/\s$ & 0 & 0 & 0  \\
         & $\eta_8 \bar K^0$ & 0 & $-1/\sx$ & $1/\sx$ & 0 & 0 & 0 \\ \hline
\end{tabular}
\end{center}
\end{table}

\begin{table}
\caption{Decomposition of $B \to PP$ amplitudes for $\Delta C = 0,~
|\Delta S| = 1$ transitions in terms of graphical contributions.}
\begin{center}
\begin{tabular}{l l c c c c c c} \hline
         & Final &  $T'$ &  $C'$ &  $P'$ &  $E'$ &  $A'$ &  $PA'$ \\
         & state &       &       &       &       &       &        \\ \hline
$B^+\to$ & $\pi^+K^0$ & 0 &  0   &   1   &   0   &   1   &   0    \\
         & $\pi^0K^+$ & $-1/\s$ & $-1/\s$ & $-1/\s$ & 0 & $-1/\s$ &
0 \\
         & $\eta_8K^+$ & $-1/\sx$ & $-1/\sx$ & $1/\sx$ & 0 & $1/\sx$ &
0 \\ \hline
$B^0\to$ & $\pi^-K^+$ & $-1$ & 0 &  $-1$ &   0   &   0   &   0    \\
         & $\pi^0K^0$ & 0 & $-1/\s$ & $1/\s$ & 0 & 0   &   0    \\
         & $\eta_8K^0$ & 0 & $-1/\sx$ & $1/\sx$ & 0 & 0  &   0    \\ \hline
$B_s\to$ & $\pi^+\pi^-$ & 0 &  0 &   0   &  $-1$ &   0   &  $-1$  \\
         & $\pi^0\pi^0$ & 0 &  0 &   0   & $1/\s$ & 0  & $1/\s$ \\
         & $K^+K^-$ & $-1$ &  0  &  $-1$ &  $-1$ &   0   &  $-1$  \\
         & $K^0 \bar K^0$ & 0 & 0 &  1   &   0   &   0   &   1    \\
         & $\pi^0 \eta_8$ & 0 & $-1/\st$ & 0 & $1/\st$ & 0 & 0    \\
         & $\eta_8\eta_8$ & 0 & $-\s/3$ & $2\s/3$ & $1/3\s$ & 0 &
$1/\s$ \\ \hline
\end{tabular}
\end{center}
\end{table}

\begin{itemize}

\item The combination $C + T$ occurs in $B^+ \to \pi^+ \pi^0$;

\item The combination $C - P$ occurs in $B_s \to \pi^0 \bar K^0$;

\item The combination $P + A$ occurs in $B^+ \to K^+ \bar K^0$;

\item The combination $P + PA$ occurs in $B^0 \to K^0 \bar K^0$;

\item The combination $E + PA$ occurs in $B^0 \to K^+ K^-$.

\end{itemize}

Similarly, for the amplitudes in Table 2:

\begin{itemize}

\item The combination $P' + T'$ occurs in $B^0 \to \pi^- K^+$;

\item The combination $C' - P'$ occurs in $B^0 \to \pi^0 K^0$;

\item The combination $P' + A'$ occurs in $B^+ \to \pi^+ K^0$;

\item The combination $P' + PA'$ occurs in $B_s \to K^0 \bar K^0$;

\item The combination $E' + PA'$ occurs in $B_s \to \pi^+ \pi^-$.

\end{itemize}

It is not possible to identify linear combinations of decay amplitudes
which depend upon the six graphical contributions separately.
\bigskip

\leftline{\bf C. Linear relations among amplitudes}
\bigskip

Since each table contains twelve decay amplitudes while there are only five
linearly independent reduced matrix elements, it must be possible to find
seven amplitude relations for each. This is indeed the case. We write
relations in terms of decay amplitudes, and then translate them into
statements about the corresponding combinations of graphical
contributions.

{\it 1. $\Delta S = 0$ processes.}
A familiar isospin relation for $B \to \pi \pi$ decays, expressing the fact
that there is just one $I = 0$ and one $I = 2$ amplitude as a result of the
form of the interaction giving rise to the decay, is [2]
\begin{equation}
\label{tri}
A(B^0 \to \pi^+ \pi^-) + \s A(B^0 \to \pi^0 \pi^0) =
\s A(B^+ \to \pi^+ \pi^0)~~~,
\end{equation}
or
\begin{equation}
-(T + P + E + PA) + (-C + P + E + PA) = -(C+T)~~~.
\end{equation}
If the rates are such that the triangle must have nonzero area, we conclude
either that there are different interactions in the $I = 0$ and $I = 2$
final states, or that there are important contributions from amplitudes
with different CKM phases (such as $T$ and $P$), or both.

There are two other triangular relations involving only pions and kaons in
the final state:
\begin{equation}
\label{kka}
\s A(B_s \to \pi^0 \bar K^0)  =
\s A(B^0 \to \pi^0 \pi^0) + A(B^0 \to K^+ K^-)~~~,
\end{equation}
\begin{equation}
i.~e.,~~~(-C + P) = (-C + P + E + PA) - (E + PA)~~~,
\end{equation}
and
\begin{equation}
\label{kkb}
A(B_s \to \pi^+ K^-) = A(B^0 \to \pi^+ \pi^-) - A(B^0 \to K^+ K^-)~~~,
\end{equation}
\begin{equation}
i.~e.,~~~-(T + P) = -(T + P + E + PA) + (E + PA)~~~.
\end{equation}
We shall anticipate a result of Sec.~III in which the $B^0 \to K^+ K^-$
amplitude may be very small if rescattering effects are negligible. In that
case each of the last two relations connects a rate for a charge state of
$B_s \to \pi \bar K$ to that for a charge state of $B^0 \to \pi \pi$. For
completeness, we include a fourth triangle relation which can be
constructed from the above three, and hence is not independent:
\begin{equation}
A(B_s \to \pi^+ K^-) + \s A(B_s \to \pi^0 \bar K^0) =
\s A(B^+ \to \pi^+ \pi^0)~~~,
\end{equation}
\begin{equation}
i.~e.,~~~-(T + P) + (-C + P) = -(C+T)~~~.
\end{equation}

In addition there are three equations relating the amplitudes for decays
involving a single $\eta_8$ to linear combinations of two other amplitudes,
and one relation between $A(B^0 \to \eta_8 \eta_8)$ and a linear
combination of three other amplitudes. These last four relations are
expected to be of limited usefulness, as we have mentioned, since the
physical $\eta$ is rather far from a pure octet. If the reader wishes to
assume that no other graphs contribute to the production of physical $\eta$
states, s/he is welcome to do so, at the risk of ignoring additional SU(3)
octet amplitudes when one $\eta$ is produced and an additional SU(3)
singlet amplitude when two $\eta$'s are produced.

{\it 2. $|\Delta S| = 1$ processes.}
The weak Hamiltonian giving rise to $\Delta C = 0$, $|\Delta S| = 1$ decays
has pieces which transform as $\Delta I = 0$ and $\Delta I = 1$. In the
decays $B \to \pi K$ there are thus {\it two} separate $I = 1/2$ amplitudes
and one $I = 3/2$ amplitude. Thus, one can write a relation among the
amplitudes for the four different charge states [3-5]:
\begin{equation}
\label{quad}
\s A(B^+ \to \pi^0 K^+) + A(B^+ \to \pi^+ K^0) =
\s A(B^0 \to \pi^0 K^0) + A(B^0 \to \pi^- K^+),
\end{equation}
or
\begin{equation}
-(T'+C'+P'+A') + (P'+A') = (-C'+P') - (T'+P')~~~.
\end{equation}
Both sides correspond to the combination $-(C' + T')$ with isospin 3/2. We
shall have a good deal more to say about this relation in what follows. It
has been the object of extensive analyses regarding the possibility of
observing direct $CP$ violation in $B \to \pi K$ decays.

The two other relations involving only pions and kaons are
\begin{equation}
\label{spia}
A(B_s \to \pi^+ \pi^-) = - \s A(B_s \to \pi^0 \pi^0)~~~,
\end{equation}
\begin{equation}
i.~e.,~~~-(E' + PA') = - (E' + PA')
\end{equation}
and
\begin{equation}
\label{spib}
A(B^0 \to \pi^- K^+) = A(B_s \to K^+ K^-) - A(B_s \to \pi^+ \pi^-)~~~,
\end{equation}
\begin{equation}
i.~e.,~~~-(T'+P') = -(T'+P'+E'+PA') + (E'+PA')~~~.
\end{equation}
The first of these follows from isospin alone; the $I = 2$ final state is
not produced. As we shall see in Sec.~III, the $B_s \to \pi \pi$ branching
ratios are expected to be extremely small if rescattering effects do not
alter the predictions of diagrams. The second relation implies that if
$(B^0 \to \pi^- K^+)$ is observed with a branching ratio of order
$10^{-5}$, as appears possible [15], at least one of the processes on the
right-hand side must be present with a similar branching ratio. Our
conjecture based on the results of Sec.~III will be that it is the process
$B_s \to K^+ K^-$.

There are also four linear relations involving decays to $\eta_8$.

{\it 3.  Relations between $\Delta S = 0$ and $|\Delta S| = 1$ processes.}
The unprimed and primed amplitudes are related, since they involve
different CKM factors but similar hadronic physics. There are two classes
of such relations. The primed non-penguin amplitudes are related to the
unprimed ones by the ratio $r_u \equiv V_{us}/V_{ud} \approx 0.23$:
\begin{equation}
\label{ckmu}
T'/T = C'/C = E'/E = A'/A = r_u~~~.
\end{equation}
The primed penguin amplitudes are related to the unprimed ones by the ratio
$r_t \equiv V_{ts}/V_{td}$, since the penguin amplitudes are dominated by
the top quark loop [16]. This quantity has a magnitude of about $5 \pm 2$.
It has a complex phase if CKM phases indeed are the source of the observed
$CP$ violation in the kaon system. Thus
\begin{equation}
\label{ckmt}
P'/P = PA'/PA = r_t~~~.
\end{equation}

It is possible to write three independent amplitude relations which involve
only $r_u$ by choosing processes that contain the combinations $C + T$, $T
- A$, and $T + E$. The relation involving $C + T$ is particularly useful
since it allows one to relate the $B^+ \to \pi^+ \pi^0$ amplitude (with
isospin 2) to the linear combination of $B \to \pi K$ amplitudes already
written above with isospin 3/2:
\begin{equation}
\label{ct}
\s A(B^+ \to \pi^0 K^+) + A(B^+ \to \pi^+ K^0) = r_u \s A(B^+ \to \pi^+
\pi^0)~~~,
\end{equation}
\begin{equation}
\label{cta}
i.~e.,~~~-(T'+C'+P'+A') + (P'+A') = - r_u(C+T)~~~.
\end{equation}
In SU(3) language, these combinations of amplitudes correspond to a pure
${\bf 27}$-plet.

A relation involving the combination $T - A$ is
\begin{equation}
\label{ta}
A(B^+ \to \pi^+ K^0) + A(B^0 \to \pi^- K^+) = r_u [A(B^+ \to K^+ \bar
K^0) + A(B_s \to \pi^+ K^-)],
\end{equation}
or
\begin{equation}
(P'+A') - (T'+P') = r_u[ (P+A) - (T+P)]~~~.
\end{equation}
A relation involving the combination $T + E$ is
\begin{equation}
\label{te}
A(B_s \to K^+ K^-) + A(B_s \to K^0 \bar K^0) = r_u [A(B^0 \to \pi^+ \pi^-)
+ A(B^0 \to K^0 \bar K^0)],
\end{equation}
or
\begin{equation}
-(T'+P'+E'+PA') + (P'+PA') = r_u[ -(T+P+E+PA) + (P + PA)]~~~.
\end{equation}
We shall discuss the likely magnitudes of terms in these relations in
Sec.~III.

The ratio $r_t$ appears in the relation between two amplitudes which
involve the combinations $P' + PA'$ or $P + PA$:
\begin{equation}
\label{ppa}
A(B_s \to K^0 \bar K^0) = r_t A(B^0 \to K^0 \bar K^0)~~~.
\end{equation}

Finally, there should be one relation which involves a mixture of penguin
and non-penguin amplitudes such as $C - P$ or $P + T$. We cannot write such
a relation in as simple a form as the others.
\bigskip

\leftline{\bf D. Specific applications to $B\to \pi \pi$ and $B \to \pi K$}
\bigskip

The amplitude of the decay to a $CP$ eigenstate, $B^0\to\pi^+\pi^-$, consists
of terms which have two different CKM phases, sometimes denoted by ``tree'' and
``penguin'' phases, $\gamma={\rm arg}(V^*_{ub}V_{ud})$ and ${-\beta=\rm arg}
(V^*_{tb}V_{td})$, respectively [17]. This affects the time-dependent $CP$
asymmetry in this process, which does not measure directly an angle ($\alpha$)
of the CKM unitarity triangle [1].

The effect of the penguin term can be eliminated if one measures, besides
the time dependent rate of $B^0\to\pi^+\pi^-$, also the (time-integrated)
rates for all possible charge states in $B\to\pi\pi$ and
$\overline{B}\to\pi\pi$. This method [2] is based on the isospin triangle
relation (\ref{tri}) and on its charge conjugate relation.

Similarly, the quadrangle isospin relation (\ref{quad}), for $B\to \pi K$,
can be used to eliminate the effect of the penguin amplitude in order to
measure $\alpha$ in the time-dependent rate of $B^0\to \pi^0 K_S$ [3-5].
For that purpose one would have to measure all the eight rates of $B\to \pi
K$ and $\overline{B} \to \pi K$. Since SU(3) relates $B\to\pi\pi$ to
$B\to\pi K$, as in Eq.~(\ref{ct}), one may try to use this symmetry to
reduce the number of necessary measurements required to determine a weak
phase. Also, final state interaction phases in these two type of processes
may be related. The triangle and quadrangle relations may be used to
measure some of these phases, which determine the magnitude of $CP$
asymmetries in charged $B$ decays. Here we will recapitulate this method,
in order to demonstrate how this may work within an SU(3) framework.

{\it 1.  Isospin in $B \to \pi \pi$.}
Eq.~(\ref{tri}) follows from a decomposition in terms of amplitudes into
final states with isospin 0 and 2 [2]:
\begin{eqnarray}
A(B^0\to\pi^+\pi^-) & = & A_2-A_0~~~, \nonumber \\
\sqrt{2}A(B^0\to\pi^0\pi^0) & = & 2A_2+A_0~~~, \nonumber \\
\sqrt{2}A(B^+\to\pi^+\pi^0) & = & 3A_2
\end{eqnarray}
In terms of graphical contributions:
\begin{eqnarray}
\label{apipi}
A_2 = -{1\over 3}(T+C) & \equiv & -a_2 e^{i\gamma}e^{i\delta_2}~~~, \nonumber\\
A_0 = {1\over 3}[(2T-C+3E)+(3P+3PA)] & \equiv &
a_{0(T)}e^{i\gamma}e^{i\delta_T} +a_{0(P)}e^{-i\beta}e^{i\delta_P}~~~.
\end{eqnarray}
On the right-hand sides the amplitudes are decomposed into terms with
different weak phases, ${\rm Arg}(V^*_{ub}V_{ud})=\gamma,~{\rm Arg}
(V^*_{tb}V_{td})=-\beta$, and different final-state-interaction phases
$\delta_2$, $\delta_T,~\delta_P$. Similar relations hold for the
charge-conjugated amplitudes, $\overline{A}$, in which one only changes the
sign of the weak phases. For convenience, let us define $\tilde{A}\equiv
e^{2i\gamma}\overline{A}$, and let us rotate all amplitudes by a phase
factor $e^{-i\gamma}e^{-i\delta_2}$. We then find:
\begin{eqnarray}
A_2 & = & \tilde{A}_2=-a_2~~~, \nonumber \\
A_0 & = & a_{0(T)}e^{i\Delta_T}-a_{0(P)}e^{i\alpha}e^{i\Delta_P}~~~,
\nonumber \\
\tilde{A}_0 & = & a_{0(T)}e^{i\Delta_T}-a_{0(P)}e^{-i\alpha}e^{i\Delta_P}~~~,
\end{eqnarray}
where $\Delta_i\equiv\delta_i-\delta_2$.

The triangle relations for $B\to\pi\pi$, Eq.~(\ref{tri}), and
$\overline{B}\to\pi\pi$ are shown in Fig.~2. The two triangles are fixed by
measurements of all six decay rates, which determine the sides of the
triangles. This determines the angle $\theta$. The coefficient of the
$\sin(\Delta mt)$ term in the time-dependent decay rate of
$B^0\to\pi^+\pi^-$ is given by $\vert A(\overline{B}^0\to\pi^+\pi^-)/
A(B^0\to\pi^+\pi^-)\vert \sin(2\alpha+\theta)$, which can then be used to
determine $\alpha$.

$CP$ violation in direct decays would be manifested by two triangles with
different shapes. This requires final-state interaction phase differences.
The final-state interaction phase difference $\Delta_P$ can be determined
from the phase of
\begin{equation}
A_0-\tilde{A}_0=-2ia_{0(P)}\sin\alpha \, e^{i\Delta_P},
\end{equation}
as shown in Fig.~2.

{\it 2. Isospin in $B \to \pi K$.}

In $B\to \pi K$ the isospin amplitudes consist of two $\Delta I=1$ ``tree''
amplitudes, $A_{1/2},~A_{3/2}$ into final states with $I=1/2,~3/2$, and an
amplitude $B_{1/2}$ of a $\Delta I=0$  operator, which is a ``tree-
penguin'' mixture. This leads to the following decomposition [3-5]:
\begin{eqnarray}
A(B^0\to\pi^- K^+) & = & A_{3/2}+A_{1/2}-B_{1/2}~~~, \nonumber \\
\sqrt{2}A(B^0\to\pi^0 K^0) & = & 2A_{3/2}-A_{1/2}+B_{1/2}~~~, \nonumber \\
A(B^+\to\pi^+ K^0) & = & A_{3/2}+A_{1/2}+B_{1/2}~~~, \nonumber \\
\sqrt{2}A(B^+\to\pi^0 K^+) & = & 2A_{3/2}-A_{1/2}-B_{1/2}~~~.
\end{eqnarray}
This implies the quadrangle relation (\ref{quad}), where both sides of this
equation express one of the diagonals of the quadrangle. A similar
quadrangle, with a common diagonal, describes the amplitudes of the
charge-conjugated processes $\overline{B}\to\pi K$. The other two diagonals
of the two quadrangles can be shown to have a common midpoint. Except for a
class of ambiguities enumerated in Ref.~[4], this suffices to specify the
shapes of both quadrangles using the eight rate measurements. The relative
orientation of the two quadrangles can then be used to eliminate the
penguin contribution from the time-dependent rate of the decay to the $CP$
eigenstate $B^0\to\pi^0 K_S$, in order to enable a measurement of $\alpha$.

{\it 3. SU(3) relations between amplitudes and strong phases in $B \to \pi
\pi$ and $B \to \pi K$.} Since the previous method involves a large number
of rate measurements, with some ambiguity in the shape of the quadrangles,
one may want to use SU(3) to relate $B\to\pi K$ to $B\to\pi\pi$. Within
SU(3) the common diagonal of the two quadrangles is given by
\begin{equation}
3A_{3/2} = -(T'+C') = r_u 3A_2 = r_u \sqrt{2}A(B^+\to\pi^+\pi^0)
\end{equation}
as observed already in (\ref{ct}). The rate of $B^+\to\pi^+\pi^0$
determines the magnitude of the common diagonal, and thus circumvents a
good deal of the uncertainty associated with the geometric construction of
the two quadrangles. Specifically, to determine this diagonal, one can make
use just of rates for charged $B$ decays, as in (\ref{ct}). The quadrangle
(\ref{quad}) and the triangle (\ref{ct}) can be combined to yield another
triangle
\begin{equation}
\label{ctnew}
\s A(B^0 \to \pi^0 K^0) + A(B^0 \to \pi^- K^+) = r_u \s A(B^+ \to \pi^+
\pi^0)~~~,
\end{equation}
\begin{equation}
\label{ctanew}
i.~e.,~~~-(C' - P') - (T' + P') = - r_u(C+T)~~~.
\end{equation}
An analysis completely analogous to that of $B\to \pi\pi$ can now be used:
the decay rates for the three processes in this triangle and their
$CP$-conjugates can be combined with the coefficient of the $\sin(\Delta
mt)$ term in the time-dependent rate for $B^0 \to \pi^0 K_S$ to determine
$\alpha$. This is the first of several examples we present in this paper
showing how SU(3) can be used to obtain information about CKM phases.

The triangle (\ref{ct}) relating the three amplitudes of charged $B$ decays
to $\pi\pi$ and $\pi K$  is interesting in its own right, since its
construction requires only self-tagging modes. This triangle is very
similar to that of $B\to \pi\pi$. One would like to find a relation between
the corresponding measurable final state phases appearing in the two
triangles. For that purpose, let us complete the analogy  by defining an
$I=1/2$ amplitude (parallel to $A_0$ in $B\to \pi\pi$) as follows:
\begin{equation}
\label{apik}
C_{1/2}\equiv -(A_{1/2}+B_{1/2})=-{1\over 3}[(T'+C'+3A')+3P']
\equiv a_{1/2(T)}e^{i\gamma}e^{i\delta_T'}-a_{1/2(P)}e^{i\delta_P'}
\end{equation}
where we used Table 2 for the expressions in terms of graphical
contributions, and ${\rm Arg}(V^*_{ub}V_{us})=\gamma,~{\rm Arg}(V^*_{tb}
V_{ts})=\pi$.

Here $\delta_{T'}$ and $\delta_{P'}$
Comparing (\ref{apik}) with (\ref{apipi}) we note that both $A_0$ and
$C_{1/2}$ consist of two terms with specific weak phases, which however
involve different graphical contributions, and thus have in general
different final state interaction phases,
$\delta_T'\not=\delta_T,~\delta_P'\not=\delta_P$. Following the arguments
which led to (31) one can similarly show that the final-state interaction
phase-difference $\Delta_{P'}\equiv
\delta_{P'}-\delta_{3/2}~(\delta_{3/2}=\delta_2)$ can be determined from
the phase of
\begin{equation}
C_{1/2}-\tilde{C}_{1/2}=2ia_{1/2(P)}\sin\gamma e^{i\Delta_{P'}},
\end{equation}
as shown in Fig.~3.

In the case $E+PA=0$, to be discussed in Sec.~III, only the diagram $P$ in
$B\to\pi\pi$ carries the ``penguin'' weak phase as in $B\to\pi K$.
Neglecting the phase due to the perturbatively calculated absorptive part
of the physical $c \overline{c}$ quark pair in the penguin diagram, which
is very small [18], one has $\delta_{P'}=\delta_P$. Thus, in the limit
$E+PA=0$, the two final state phase differences in $B\to\pi\pi$ and in
$B\to\pi K$ are equal, $\Delta_{P'}=\Delta_P$.
\bigskip

\leftline{\bf E. Isospin analysis of $B \to \pi \bar D$}
\bigskip

Although we have concentrated on decays of $B$ mesons to pairs of light
pseudoscalars, some useful evidence regarding final-state interactions can
be obtained from the processes $B \to \pi \bar D$. The transition $\bar b
\to \bar d u \bar c$ has $\Delta I = \Delta I_3 = 1$ and hence there are
unique amplitudes for $I = 1/2$ and $I = 3/2$ final states. In terms of
these, the physical decay amplitudes may be written
\begin{equation}
A(B^+ \to \pi^+ \bar D^0) = A_{3/2}~~~,
\end{equation}
\begin{equation}
A(B^0 \to \pi^+ D^-) = \frac{1}{3}A_{3/2} + \frac{2}{3}A_{1/2}~~~,
\end{equation}
\begin{equation}
A(B^0 \to \pi^0 \bar D^0) = \s(A_{3/2} - A_{1/2})/3~~~.
\end{equation}
These equations are of the same form as those employed to conclude that
significant final-state interactions occur in $D \to \pi \bar K$ decays.
They imply the triangle relation
\begin{equation}
A(B^0 \to \pi^+ D^-) + \s A(B^0 \to \pi^0 \bar D^0) =
A(B^+ \to \pi^+ \bar D^0)~~~.
\end{equation}
Experimental data [7] exist for the two processes involving charged pions,
but only an upper limit exists at present for $B^0 \to \pi^0 \bar D^0$.
These results lead to an upper limit on the phase shift difference [19],
$\cos(\delta_{1/2} - \delta_{3/2}) > 0.82$ (90\% c.l.), or $|\delta_{1/2} -
\delta_{3/2}| < 35^{\circ}$. With improved data, one may be able to tell
whether the triangle has nonzero area. Since these decays are all expected
to be governed by the same CKM factor, nonzero area for the triangle would
be unambiguous evidence for a difference in final-state phases between the
$I = 1/2$ and $I = 3/2$ amplitudes. A similar approach [19] failed to
detect any phase shift differences between $I = 1/2$ and $I = 3/2$
amplitudes in the decays $B \to \pi \bar D^*$ and $B \to \rho \bar D$.
\bigskip

\centerline{\bf III. NEGLECT OF CERTAIN DIAGRAMS}
\bigskip

\leftline{\bf A. Linear relations among reduced amplitudes}
\bigskip
The diagrams denoted by $E,~A,~PA$ involve contributions to amplitudes
which should behave as $f_B/m_B$ in comparison with those from the diagrams
$T,~C$, and $P$ (and similarly for their primed counterparts). This
suppression is due to the smallness of the $B$ meson wave function at the
origin, and it should remain valid unless rescattering effects are
important. Such rescatterings indeed could be responsible for certain
decays of charmed particles (such as $D^0 \to \bar K^0 \phi$ [20]), but
should be less important for the higher-energy $B$ decays. In addition the
diagrams $E$ and $A$ are also helicity suppressed by a factor
$m_{u,d,s}/m_B$ since the $B$ mesons are pseudoscalars. We shall
investigate in the present section the consequences of assuming that only
$T,~C,~P$ and the corresponding primed quantities are nonvanishing.

The relations for reduced matrix elements in SU(3) entailed by this
assumption are surprisingly simple.  The singlet and the octet (${\bf 8}_1$)
arising from the ${\bf 3^*}$ operator in the weak hamiltonian become
related to one another, while the {\bf 27}-plet amplitude and the octet
(${\bf 8}_3$) which arises from the ${\bf 15^*}$ operator become related:
\begin{equation}
\label{rel}
\{8_1\} = - \sqrt{5}\{1\}/4~~,~~~\{8_3\} = \{27\}/4~~~.
\end{equation}
The amplitude $\{8_2\}$ which arises from the ${\bf 6}$ operator remains
unconstrained.
\bigskip

\leftline{\bf B. Relations among decay amplitudes}
\bigskip

There are now three independent SU(3) amplitudes for $\Delta S = 0$
transitions expressed in terms of the three independent graphical
contributions $T,~C$, and $P$, and three for $|\Delta S| = 1$ transitions
expressed in terms of $T',~C'$, and $P'$. The relations (\ref{ckmu}) and
(\ref{ckmt}) between these two sets noted in Sec.~II continue to hold.

When the $E,~A,~PA$ diagrams and their primed counterparts are neglected,
certain decays become forbidden:
\begin{equation}
\Gamma(B^0 \to K^+ K^-) = 0~~~,
\end{equation}
\begin{equation}
\Gamma(B_s \to \pi^+ \pi^-) = \Gamma(B_s \to \pi^0 \pi^0) = 0~~~.
\end{equation}
This leaves 13 nonzero amplitudes involving pions and kaons, as shown in
Table 3. (In this Table, the $\s(B^+\to\pi^+\pi^0)$ under the $-(T+C)$
column means that $A(B^+\to\pi^+\pi^0) = -(T+C)/\s$, and similarly for
other entries.)

\begin{table}
\caption{The 13 decay amplitudes in terms of the 8 graphical combinations.}
\begin{center}
\begin{tabular}{l l c c c c c c} \hline
$~~~-(T+C)$ & $~~~-(C-P)$         & $-(T+P)$     &
  $(P)$               & \\ \hline
$\s(B^+\to\pi^+\pi^0)$  & $\s(B^0\to\pi^0\pi^0)$ & $B^0\to\pi^+\pi^-$ &
$B^+\to K^+ \Kbar^0$ & \\
                    & $\s(B_s\to\pi^0 \Kbar^0)$ & $B_s\to\pi^+K^-$   &
$B^0\to K^0 \Kbar^0$  &  \\ \hline
$~-(T'+C'+P')$ & $~~~-(C'-P')$ & $-(T'+P')$     &
  $(P')$               & \\ \hline
$\s(B^+\to\pi^0 K^+)$  & $\s(B^0\to\pi^0 K^0)$ & $B^0\to\pi^- K^+$ &
$B^+\to\pi^+ K^0$ & \\
                       &                   & $B_s\to K^- K^+$  &
$B_s\to K^0 \Kbar^0$  &  \\ \hline
\end{tabular}
\end{center}
\end{table}

Given the relations (\ref{ckmu}) and (\ref{ckmt}), these 13 amplitudes can
be expressed in terms of 3 independent quantities, leading to 10 relations
among the amplitudes. For $\Delta S=0$ processes, the triangle relation
(\ref{tri}) for $B \to \pi \pi$ is not simplified by the assumptions of the
present Section. However, the two triangle relations (\ref{kka}) and
(\ref{kkb}) of the exact treatment become two rate relations:
\begin{equation}
\Gamma(B_s \to \pi^0 \bar K^0) = \Gamma(B^0 \to \pi^0 \pi^0)~~~,
\end{equation}
\begin{equation}
\Gamma(B_s \to \pi^+ K^-) = \Gamma(B^0 \to \pi^+ \pi^-)~~~.
\end{equation}
As for $\vert \Delta S \vert = 1$ processes, the quadrangle relation
(\ref{quad}) for $B \to \pi K$ is unchanged. However, as mentioned in
Sec.~II D, in the context of SU(3) it is more convenient to think of this
quadrangle relation as {\it not} independent, since it can be expressed in
terms of two triangle relations. As a result of the vanishing of the $B_s
\to \pi^+ \pi^-$ transition, the triangle relation (\ref{spib}) becomes a
rate relation:
\begin{equation}
\Gamma(B^0 \to \pi^- K^+) = \Gamma(B_s \to K^- K^+)~~~.
\end{equation}

There is one new relation between amplitudes which are pure $P$:
\begin{equation}
A(B^+ \to K^+ \bar K^0) = A(B^0 \to K^0 \bar K^0)
\end{equation}
and one new relation involving pure $P'$:
\begin{equation}
A(B^+ \to \pi^+ K^0) = A(B_s \to K^0 \bar K^0)~~~.
\end{equation}

Since above we have listed 6 relations [counting the triangle relation
(\ref{tri})], there should be 4 relations between $\Delta S=0$ and $\vert
\Delta S \vert = 1$ processes. Three of these, the two triangle relations
(\ref{ct}) and (\ref{ctnew}) and the rate relation (\ref{ppa}), are
unchanged from the exact treatment. [The $\pi K$ quadrangle relation
(\ref{quad}) is simply related to the two triangle relations (\ref{ct}) and
(\ref{ctnew}).] Finally, given the assumptions of this Section, the two
quadrangle relations (\ref{ta}) and (\ref{te}) become equivalent, and can
be written in terms of decays of $B^0$ and $B^+$ only:
\begin{equation}
\label{tanew}
A(B^+ \to \pi^+ K^0) + A(B^0 \to \pi^- K^+) = r_u [A(B^+ \to K^+ \bar
K^0) + A(B^0 \to \pi^+ \pi^-)],
\end{equation}
\begin{equation}
i.~e.,~~~(P') - (T'+P') = r_u[ (P) - (T+P)]~~~.
\end{equation}
Note that the left-hand side of this relation, and others like it, is
likely to involve the cancellation of two nearly equal amplitudes if $P'$
is the dominant effect in $|\Delta S| = 1$ transitions, as we expect to be
likely (see below).

To sum up, the assumption of ignoring the diagrams $E,~A,~PA$ and their
corresponding primed quantities leads to 10 relations among $B$ decay
amplitudes: 6 rate relations, 3 triangle relations and one quadrangle
relation. These will be very useful in extracting both weak and strong
phase information from $B$ decays.

Finally, there is an additional point: the penguin contributions in $B \to
\pi \pi$ and $B \to \pi K$ are now related to one another since the
amplitude $PA$ is no longer present in $B \to \pi \pi$. Thus, we expect the
strong phase shift difference $\delta_P - \delta_2$ in $B \to \pi \pi$ to
be equal to the difference $\delta_P - \delta_{3/2}$ in $B \to \pi K$. We
showed in Sec.~II D how to measure these differences.
\bigskip

\leftline{\bf C. Measuring the angle $\gamma$}
\bigskip

Neglecting $A'$ in (\ref{cta}), the triangle relation (\ref{ct}) and its
charge-conjugate can be used to measure the angle $\gamma$ [21]. This
follows from the fact that the amplitude $-(C+T) = \sqrt{2} A(B^+ \to
\pi^+\pi^0)$ has the tree weak phase $\gamma$, whereas the amplitude $P' =
A(B^+\to\pi^+ K^0)$ has the penguin phase $\pi$. The two triangles for
$B^+$ and $B^-$ decays are shown in Fig.~4, with the notation defined in
the caption. We have drawn the figure such that the strong phase $e^{i
\delta_P}$ lies along the horizontal axis. The strong phase $\delta_2$ was
discussed in Sec.~II D 1.

The measurements of the four independent rates for $B^+ \to \pi^0 K^+$,
$B^- \to \pi^0 K^-$, $B^+ \to \pi^+ K^0$, and $B^+ \to \pi^+ \pi^0$ can
determine $\gamma$. If $A'$ can be neglected, the rates for $B^+ \to \pi^+
K^0$ and $B^- \to \pi^- \bar K^0$ should be equal. We expect $\Gamma(B^+
\to \pi^+ \pi^0) = \Gamma(B^- \to \pi^- \pi^0)$ in any case, as noted
earlier. The triangle for $B^-$ decays can also be flipped about its
horizontal axis, leading to a two-fold ambiguity.

If $\delta_P - \delta_2 = 0$, we will not observe a $CP$-violating
difference between the rates for $B^+ \to \pi^0 K^+$ and $B^- \to \pi^0
K^-$. Nevertheless, accepting the standard model CKM mechanism for $CP$
violation in the kaon system, we know that $\gamma \ne 0$, which selects
the ``flipped'' solution as shown in the lower figure.
\bigskip

\leftline{\bf D. Measuring $\beta$, $\gamma$, and the strong phase shifts}
\bigskip

The relations of Sec.~III B can be also used to measure the weak phases
$\beta$ and $\gamma$, as well as strong phase shifts [22]. Of the 10
relations, there are 3 triangle relations, which can be written
schematically as
\begin{equation}
(T+C)=(C-P)+(T+P)~,
\end{equation}
\begin{equation}
(T+C)=(C'-P')/r_u + (T'+P')/r_u~,
\end{equation}
\begin{equation}
(T+C) = (T'+C'+P')/r_u - (P')/r_u~.
\end{equation}
By SU(3) symmetry the strong phases for the primed graphs are the same as
the unprimed ones. We thus have for the three relations
\begin{equation}
3 a_2 e^{i\gamma} e^{i\delta_2}
=(A_C e^{i\gamma}e^{i\delta_C}-A_P e^{-i\beta}e^{i\delta_P})
+(A_T e^{i\gamma}e^{i\delta_T}+A_P e^{-i\beta}e^{i\delta_P})~~~,
\end{equation}
\begin{equation}
3 a_2 e^{i\gamma} e^{i\delta_2}
=(A_C e^{i\gamma}e^{i\delta_C}+A_{P'} e^{i\delta_P}/r_u)
+(A_T e^{i\gamma}e^{i\delta_T}-A_{P'} e^{i\delta_P}/r_u)~~~,
\end{equation}
\begin{equation}
3 a_2 e^{i\gamma} e^{i\delta_2}
=(A_T e^{i\gamma}e^{i\delta_T} + A_C e^{i\gamma}e^{i\delta_C}
-A_{P'} e^{i\delta_P}/r_u)
+A_{P'} e^{i\delta_P}/r_u~~~,
\end{equation}
where $a_2$ [introduced in Eq.~(\ref{apipi})] and the quantities $A_T$,
$A_C$, $A_P$ and $A_{P'}$ are real and positive. Multiplying through on
both sides by $\exp(-i\gamma-i\delta_2)$ gives
\begin{equation}
\label{taa}
3a_2
=(A_C e^{i\Delta_C}+A_P e^{i\alpha}e^{i\Delta_P})
+(A_T e^{i\Delta_T}-A_P e^{i\alpha}e^{i\Delta_P})~~~,
\end{equation}
\begin{equation}
\label{tb}
3a_2
=(A_C e^{i\Delta_C} + A_{P'} e^{-i\gamma}e^{i\Delta_P}/r_u)
+(A_T e^{i\Delta_T} - A_{P'} e^{-i\gamma}e^{i\Delta_P}/r_u)~~~,
\end{equation}
\begin{equation}
\label{tc}
3a_2
=(A_T e^{i\Delta_T} + A_C e^{i\Delta_C}
- A_{P'} e^{-i\gamma}e^{i\Delta_P}/r_u)
+ A_{P'} e^{-i\gamma}e^{i\Delta_P}/r_u)~~~,
\end{equation}
where we again define $\Delta_i\equiv\delta_i-\delta_2$, and note that
$-(\beta + \gamma) = \alpha - \pi$.

Consider the two triangle relations in Eqs.~(\ref{tb}) and (\ref{tc}).
Implicit there is the relation
\begin{equation}
\label{cttri}
3a_2 = T + C = A_T e^{i\Delta_T} + A_C e^{i\Delta_C}~~~.
\end{equation}
Thus not only do both these triangles share a common base but they also
share a common subtriangle with sides $T + C,~C,~T$ as shown in Fig.~5(a).
Furthermore, this subtriangle is completely determined, up to a four-fold
ambiguity, by the two triangles in Eqs.~(\ref{tb}) and (\ref{tc}). In other
words, if we measure the five rates for $B^0 \to \pi^-K^+$ (giving $|T +
P'/r_u|$), $B^0 \to \pi^0 K^0$ (giving $|C - P'/r_u|$), $B^+ \to \pi^0 K^+$
(giving $|T + C + P'/r_u|$), $B^0 \to \pi^+ K^0$ (giving $|P'/r_u|$), and
$B^0 \to \pi^+\pi^0$ (giving $|T + C|$), we can determine $|T|$, $|C|$,
$\Delta_C$, $\Delta_T$, and $\Delta_P - \gamma$. If we also measure the
$CP$ conjugate processes then we can also separately determine $\Delta_P$
and $\gamma$.

Now consider the triangle relations in Eqs.~(\ref{taa}) and (\ref{tb}).
Since the magnitudes of the penguin diagrams $P$ and $P'$ can be measured
by measuring rates, the sub-triangle in Eq.~(\ref{cttri}) (which still
holds) is completely determined up to an eight-fold ambiguity. This
eight-fold ambiguity corresponds to the two intersections of the circles
drawn from the vertices of the triangles as seen in Fig.~5(b), in addition
to the four-fold ambiguity caused by the possibility of reflecting the
triangles about their bases. Thus by measuring 7 rates we can extract, in
addition to the parameters mentioned above, the angle $\Delta_P-\alpha$.
Thus we can determine $\beta=\pi-\alpha-\gamma$. By considering the $CP$
conjugate processes we can determine $\Delta_P$ and $\alpha$ separately.

An identical construction to that in the previous paragraph holds for
relations  (\ref{tb}) and (\ref{tc}). This provides an independent way to
measure the same quantities and is likely to be of great help in evaluating
the size of SU(3)-breaking effects [22].

In Figs.~5 more realistic proportions would be $|P|,~|C| < |T| < |P'/r_u|$.
This hierarchy can reduce the possibility of discrete ambiguities.  Thus,
for example, in Fig.~5(a) two sides of each triangle with base $C + T$ will
be of order $|P'/r_u|$. One of the two choices of relative orientation of
the two triangles will imply that $|C|$ and $|T|$ are each of order
$|P'/r_u|$, violating this hierarchy. Thus only a {\it two-fold} ambiguity
will remain, corresponding to reflection of {\it each} triangle about the
base $C + T$.

\bigskip

\leftline{\bf E. Results of further specialization}
\bigskip

The relative magnitude of penguin effects in the decays $B \to \pi \pi$ can
be estimated either by direct reference to various charge states in $B \to
\pi \pi$ [2], or with the help of SU(3) and some auxiliary assumptions by
reference to the process $B^0 \to \pi^- K^+$ [11]. It appears that some
combination of the decays $B^0 \to \pi^+ \pi^-$ and $B^0 \to \pi^- K^+$ has
been observed, with the most likely mixture being approximately equal
amounts of each [15]. As a result of the relations (\ref{ckmu}) and
(\ref{ckmt}) one then concludes that $T$ and $C$ are likely to dominate the
$\Delta S = 0$ transitions, while $P'$ is likely to dominate the $|\Delta
S| = 1$ transitions. For reference, we quote some results of assuming this
to be so.

{\it 1. $B \to \pi \pi$ without penguins.} If the triangle formed by the
complex amplitudes in (\ref{tri}) has non-zero area and penguin
contributions are known to be small, there must be final state phase
differences between $I = 0$ and $I = 2$ amplitudes. Direct $CP$ violation
in rates is not observable but the usual measurement of a $CP$-violating
difference between the time-integrated rates for $B^0|_{t=0} \to \pi^+
\pi^-$ and $\bar B^0|_{t=0} \to \pi^+ \pi^-$ measures the angle $\alpha$ in
the unitarity triangle. Here $\alpha + \beta + \gamma = \pi$, with $\beta =
{\rm Arg}~V^*_{td} V_{tb}$ and $\gamma = {\rm Arg}~ V^*_{ub} V_{ud}$.

If the magnitude $|P|$ is {\it measured} to be small in comparison with $T$
and $C$ (by the study of $B^+ \to K^+ \bar K^0$ and $B^0 \to K^0 \bar K^0$
as mentioned below), and all the other graphs give negligible
contributions, one can draw the triangle of Eq.~(\ref{tri}) as shown in
Fig.~6, with a circle of error around one vertex corresponding to the
uncertainty in the phase of $P$. (This assumes we have not yet measured
that phase using methods mentioned earlier.) If this circle is small
enough, we can obtain approximate information on the relative strong phases
of the amplitudes $C$ and $T$, and hence on the phase difference $\delta_2
- \delta_0$ in $B \to \pi \pi$.

{\it 2. $B \to \pi K$ with penguins alone.} All rates are related [9];
those with charged pions are twice those with neutral ones. The quadrangles
in Fig.~3 have zero area because their common diagonal $d_1$ receives no
contributions from penguin amplitudes and hence vanishes. As mentioned in
Sec.~II, one expects to be able to tell directly from the $B^+ \to \pi^+
\pi^0$ rate how large this diagonal actually is. If the quadrangles have
zero area, the $CP$-violating difference between the time-integrated rates
for $B^0|_{t= 0} \to \pi^0 K_S$ and $\bar B^0|_{t=0} \to \pi^0 K_S$
measures the angle $\beta$ in the unitarity triangle.

{\it 3. Other $B \to PP$ rate predictions.} If $P$ is negligible in
comparison with $C$ and $T$, the processes $B^+ \to K^+ \bar K^0$ and $B^0
\to K^0 \bar K^0$ have negligible rates in comparison with $B \to \pi \pi$.
(We have already argued that $B^0 \to K^+ K^-$ is likely to be small.) For
$|\Delta S| = 1$ transitions, there appear to be no special rate
predictions beyond those for $B \to \pi K$ which follow from the assumption
of $P'$ dominance. (We exclude processes involving $\eta_8$'s from this
discussion, as usual.)

{\it 4. Neglect of color-suppressed diagrams.} If both $P$ and $C$ are
negligible in comparison with $T$ in $\Delta S = 0~B \to PP$ decays, the
$B^0 \to \pi^0 \pi^0$ decay does not occur, and the triangle relation
(\ref{tri}) becomes a relation between two amplitudes, entailing a rate
prediction $\Gamma(B^+ \to \pi^+ \pi^0) = \Gamma(B^0 \to \pi^+ \pi^-) /2$.

If the color-suppressed amplitude $C'$ can be neglected in $B \to \pi K$ in
comparison to $T'$ and $P'$, one obtains the rate relations $2 \Gamma(B^0
\to \pi^0 K^0) = \Gamma(B^+ \to \pi^+ K^0)$ and $2 \Gamma(B^+ \to \pi^0
K^+) = \Gamma(B^0 \to \pi^- K^+)$. This may help in extracting the weak
phase from the time-dependence of $B^0 \to \pi^0 K_S$.

In decays to charmed final states, the absence of a color-suppressed
contribution would lead to equal rates for $B^+ \to \pi^+ \bar D^0$ and
$B^0 \to \pi^+ D^-$ and a suppression of $B^0 \to \pi^0 \bar D^0$. It will
be interesting to watch as the data on these processes improve, to learn
the actual suppression factor. A similar suppression is expected in $B \to
K D$. The magnitude of color suppression in this class of processes would
become crucial for a measurement of the unitarity triangle angle $\gamma$
from the rates of the self-tagged modes $B^+ \to K^+ D^0$, $B^+ \to K^+
\bar D^0$, and $B^+ \to K^+ D_{CP}$, where $D_{CP}$ denotes a $CP$
eigenstate [23]. Too strong a suppression of $B^+ \to K^+ D^0$, which only
involves a color-suppressed and an annihilation diagram, would presumably
make this method unfeasible.
\bigskip

\centerline{\bf IV. RESULTS FOR PHYSICAL $\eta$ AND $\eta'$}
\bigskip

The physical $\eta$ and $\eta'$ appear to be octet-singlet mixtures:
\begin{equation}
\eta = \eta_8 \cos \phi - \eta_1 \sin \phi~~~,~~
\eta' = \eta_8 \sin \phi + \eta_1 \cos \phi~~~,
\end{equation}
where $\eta_8 \equiv (2 s \bar s - u \bar u - d \bar d)/\sx$ and
$\eta_1 \equiv (u \bar u + d \bar d + s \bar s)/\st$.

For a mixing angle of $\phi = 19.5^{\circ} = \sin^{-1}(1/3)$, close to one
obtained in a recent analysis [14], the physical $\eta$ and $\eta'$ can be
represented approximately as
\begin{equation}
\label{physeta}
\eta = (s \bar s - u \bar u - d \bar d)/\st~~~,~~
\eta' = (u \bar u + d \bar d + 2 s \bar s)/\sx~~~.
\end{equation}
We shall calculate amplitudes for production of these states in terms
purely of the graphical contributions of Figs.~1(a)-(c), neglecting the
small terms associated with exchange, annihilation, or penguin
annihilation. We also neglect graphs in which one or two final-state
particles are connected to the rest of the diagram by gluons (or vacuum
quantum numbers) alone. It would not make sense to neglect such graphs
while still continuing, for example, to take account of the disconnected
penguin annihilation graph of Fig.~1(f). As mentioned in Sec.~II, a full
SU(3) analysis would have to take account of all types of disconnected
graphs as a result of the singlet components of $\eta$ and $\eta'$. The
validity of the neglect of disconnected graphs in charmed-particle decays
involving $\eta$ and $\eta'$ has been discussed by Lipkin [24].

The results are shown in Tables 4 and 5.

\begin{table}
\caption{Decomposition of $B \to PP$ amplitudes involving $\eta$ and
$\eta'$ for $\Delta C = \Delta S = 0$ transitions in terms of graphical
contributions. Here $\eta$ and $\eta'$ are defined as in
Eq.~(\protect\ref{physeta}).}
\begin{center}
\begin{tabular}{l l c c c} \hline
         & Final         &    $T$    &    $C$    &    $P$    \\
         & state         &           &           &           \\ \hline
$B^+\to$ & $\pi^+\eta$   & $-1/\st$  & $-1/\st$  & $-2/\st$  \\
         & $\pi^+\eta'$  &  $1/\sx$  &  $1/\sx$  &  $2/\sx$  \\ \hline
$B^0\to$ & $\pi^0\eta$   &     0     &     0     & $-2/\sx$  \\
         & $\pi^0\eta'$  &     0     &     0     &  $1/\st$  \\
         & $\eta\eta$    &     0     &  $\s/3$   &  $\s/3$   \\
         & $\eta\eta'$   &     0     & $-\s/3$   & $-\s/3$   \\
         & $\eta'\eta'$  &     0     &  $\s/6$   &  $\s/6$   \\ \hline
$B_s\to$ & $\eta \bar K^0$ &   0     & $-1/\st$  &     0     \\
         & $\eta' \bar K^0$ &  0     &  $1/\sx$  &  $3/\sx$  \\ \hline
\end{tabular}
\end{center}
\end{table}

\begin{table}
\caption{Decomposition of $B \to PP$ amplitudes for $\Delta C = 0,~
|\Delta S| = 1$ transitions involving $\eta$ and $\eta'$ as defined in
(\protect\ref{physeta}) in terms of graphical contributions.}
\begin{center}
\begin{tabular}{l l c c c} \hline
         & Final       &     $T'$    &     $C'$    &     $P'$    \\
         & state       &             &             &             \\ \hline
$B^+\to$ & $\eta K^+$  &  $-1/\st$   &  $-1/\st$   &      0      \\
         & $\eta' K^+$ &   $1/\sx$   &   $1/\sx$   &   $3/\sx$   \\ \hline
$B^0\to$ & $\eta K^0$  &      0      &  $-1/\st$   &      0      \\
         & $\eta' K^0$ &      0      &   $1/\sx$   &   $3/\sx$   \\ \hline
$B_s\to$ & $\pi^0 \eta$ &     0      &  $-1/\sx$   &      0      \\
         & $\pi^0 \eta'$ &    0      &   $1/\st$   &      0      \\
         & $\eta\eta$  &      0      &  $-\s/3$    &   $\s/3$    \\
         & $\eta\eta'$ &      0      &  $-1/3\s$   &   $\s/3$    \\
         & $\eta'\eta'$ &     0      &  $-\s/3$    &   $2\s/3$   \\ \hline
\end{tabular}
\end{center}
\end{table}

An interesting feature of these results is the absence of the $P$
contribution to $B_s \to \eta \bar K^0$ and the $P'$ contribution to $B^0
\to K^0 \eta$. The first result says that $B_s \to \eta \bar K^0$ may be a
good probe of the color-suppressed contribution $C$. This can be tested by
comparison with the value of $C$ as extracted using $B\to \pi\pi$, $\pi K$
and $KK$ (Sec.~III D). The second result says that the decay $B^0 \to K^0
\eta$ may be considerably suppressed in comparison with $B^0 \to K^0
\pi^0$. The suppression of the $P'$ contribution is complete for the
particular mixing scheme (\ref{physeta}), but is also considerable [25] for
a mixing angle of $\phi \approx 10^{\circ}$ such that $\eta \approx (\s s
\bar s - u \bar u - d \bar d)/2$.
\bigskip

\centerline{\bf V. SU(3)-BREAKING EFFECTS}
\bigskip

As pointed out in Ref.~[9], there appear to be important SU(3)-breaking
effects in charmed meson decays. One expects [26] $\Gamma(D^0 \to K^+ K^-)
/\Gamma(D^0 \to \pi^+ \pi^-) = 1$, but this ratio appears to be [6,27] $2.6
\pm 0.4$. It is possible to take at least partial account of such effects
in the case of $B$ decays. Some of them are expected to be independent of
the mass of the decaying quark and some are expected to decrease with
increasing quark mass. The independent ways we have described of extracting
the same strong and weak phases with SU(3) relations provide a way to
measure the size of SU(3) breaking effects.
\bigskip

\leftline{\bf A. Meson decay constants}
\bigskip

In the description of decays via factorization, a charged weak current can
materialize either into a pion, with decay constant $f_\pi = 132$ MeV, or
into a kaon, with decay constant $f_K = 160$ MeV. Through Fierz identities
it is sometimes assumed that neutral quark-antiquark combinations emerging
from a weak vertex materialize into neutral pseudoscalar mesons with
corresponding decay constants, though this assumption is on shakier ground.

Since $f_K/f_\pi \approx 1.2$, one can expect this effect alone to
contribute to deviations in decay rates by more than 40\% from the naive
SU(3) expectation, independently of the mass of the decaying particle.
Such effects have been taken into account in Ref.~[11] in relating penguin
effects in $ B \to \pi \pi$ to the corresponding ones in $B \to \pi K$.
\bigskip

\leftline{\bf B. Form factors and hadronization}
\bigskip

The rates for processes like $D^0 \to K^+ K^-$ and $D^0 \to \pi^+ \pi^-$,
if calculated using factorization, depend not only on meson decay constants
but also on form factors for $D \to K$ and $D \to \pi$. If $f^+_{D \to
K}(0)/f^+_{D \to \pi}(0)> 1.2$, one can understand why the rates for $D^0
\to K^+ K^-$ and $D^0 \to \pi^+ \pi^-$ are so different. The cooperation of
two different SU(3)-breaking effects (decay constants and form factors) in
this case originates in the presence of two different tree subprocesses ($c
\to d u \bar d$ and $c \to s u \bar s$). No similar case arises in $B$
decays, and therefore SU(3) breaking is generally expected to be smaller.

Some evidence that the corresponding ratio $f^+_{B \to K}(0)/f^+_{B \to
\pi}(0)$ exceeds 1, and could be of the order of $1.1 \pm 0.1$, comes from
a recent QCD sum rule calculation [28]. In Ref.~[11] this ratio of form
factors was assumed equal to 1. This ratio may be relevant to the ratio of
$b \to s$ and $b \to d$ penguin contributions, at least for the decays of
nonstrange $B$'s.

In SU(3) we assume that the probability of producing an extra $s \bar s$
pair from the vacuum equals the probability for $u \bar u$ and $d \bar d$
production. For the heavy $B$ meson this may be a good approximation.
\bigskip

\leftline{\bf C. Specific applications}
\bigskip

The SU(3) relations between $\Delta S = 0$ and $|\Delta S| = 1$ transitions
are of two types, Eqs.~(\ref{ckmu}) and (\ref{ckmt}). For $\Delta S = 0$
processes, the dominant effects are expected to be $T$ and $C$, while for
$|\Delta S| = 1$ we expect $P'$ to dominate. The small admixture of $P$ in
$\Delta S = 0$ transitions is estimated from $P'$ using (\ref{ckmt}), while
the small contributions of $T'$ and $C'$ to $|\Delta S| = 1$ transitions
are estimated using (\ref{ckmu}). In both of these cases, since SU(3) is
only used in order to estimate the magnitude of the smaller amplitude, the
effects of SU(3) breaking will be suppressed by a factor of 4 to 5 in any
given decay.

When we come to relations such as (\ref{ct}) in which two large amplitudes
nearly cancel, SU(3) breaking is more important. Here, by referring to the
graphs which give rise to the amplitudes $C$ and $T$, and assuming
factorization to govern their behavior, we can expect the main effect of
SU(3) breaking to involve the ratio $f_K/f_\pi$. We then expect (\ref{ct})
to be replaced by
\begin{equation}
\label{ctb}
\s A(B^+ \to \pi^0 K^+) + A(B^+ \to \pi^+ K^0) = (f_K/f_\pi) r_u \s A(B^+
\to \pi^+ \pi^0)~~~.
\end{equation}

It is more difficult to estimate the effects of SU(3) breaking on an
equation such as (\ref{ppa}) which involves the ratio $r_t$ in
(\ref{ckmt}). Both form factor and hadronization effects enter into
corrections to this relation.
\bigskip

\centerline{\bf VI. SUMMARY AND EXPERIMENTAL PROSPECTS}
\bigskip

We have examined the decays of $B$ mesons to two light pseudoscalars within
the context of SU(3), looking for amplitude relations, simplifications, and
help in sorting out the physics of the $B \to \pi K$ system. While the
SU(3) decompositions we obtain are not new, we have found a number of
simple linear relations among amplitudes whose validity tests assumptions
at various levels of generality.

An SU(3) analysis without further simplifying assumptions leads to several
rate predictions, a number of triangle relations among amplitudes, and one
very useful relation (\ref{ct}) between the amplitude for $B^+ \to \pi^+
\pi^0$ and the $I = 3/2$ amplitude in $B \to \pi K$. This last relation can
be used in several different ways, including the specification of relative
phases of various $B \to \pi K$ amplitudes and the substitution of a
measurement of the $B^+ \to \pi^+ \pi^0$ rate for a measurement of the
rates for $B^0 \to \pi^- K^+$ and the charge-conjugate process in sorting
out CKM phases. We have also shown how to extract strong final-state phase
differences $\delta_P - \delta_2$ or $\delta_{P'} - \delta_{3/2}$ between
the penguin amplitude and the $I = 2$ amplitude in $B \to \pi \pi$ or the
$I = 3/2$ amplitude in $B \to \pi K$.

Additional predictions are obtained if one is prepared to neglect certain
contributions in a manner motivated by a graphical SU(3) language. The
neglect of these contributions is predicated on the relative unimportance
of strong rescattering effects. These predictions frequently convert
triangle relations to relations regarding rates, since in a number of cases
they imply that one side of a triangle has vanishing length. One
application of these relations is that measurements of the rates for $B^+$
to $\pi^+ \pi^0,~\pi^+ K^0$, and $\pi^0 K^+$ and the charge-conjugate
processes can be used to determine  the weak CKM phase $\gamma$. With
measurements of the remaining rates for $B$ decays to $\pi \pi,~\pi K$, and
$K \bar K$, one can obtain the CKM phases $\gamma$ and $\alpha$ and all the
relevant differences of strong phase shifts.

In the more general case, when all amplitudes are considered, we can learn
about final-state phase shift differences from the decays $B \to \pi \bar
D$, which involve a single CKM factor. If such phase shift differences were
small, we would expect them to be even smaller in the decays to pairs of
light pseudoscalars in which more energy is released. One relies on the
presence of such phase shift differences in order to be able to detect
direct $CP$ violation in such processes as self-tagging $B \to \pi K$
decays.

As we have mentioned earlier, some combination of the decays $B^0 \to \pi^+
\pi^-$ and $B^0 \to \pi^- K^+$ has been observed [15], with a total
branching ratio of about $2 \times 10^{-5}$. If this consists of equal
amounts of $\pi \pi$ and $\pi K$, and if the $\pi K$ decays are indeed
dominated by a $\Delta I = 0$ transition as occurs in a penguin graphs,
{\it all} the charge states of $B \to \pi K$ should be observable at levels
of $10^{-5}$ for charged pions or half that for neutral pions. Similarly,
if color-unsuppressed tree diagrams dominate the $B \to \pi^+ \pi^-$
process, and if it occurs at a level of $10^{-5}$, one should see $B^+ \to
\pi^+ \pi^0$ at a level of half that. Once these signals are observed,
refinements of rate information will be able to test for the presence of
subdominant contributions. The next step would be to look for processes
which we predict to be suppressed; searches at branching ratios down to
$10^{-7}$ would be able to provide information on amplitudes at the 10\%
level and could be of great help in sorting out prospects for observing
signals of $CP$ violation.
\bigskip

\centerline{\bf ACKNOWLEDGMENTS}
\bigskip

We thank B. Blok, H. Lipkin, and L. Wolfenstein for fruitful discussions.
M. Gronau and J. Rosner respectively wish to acknowledge the hospitality of
the Universit\'e de Montr\'eal and the Technion during parts of this
investigation. This work was supported in part by the United States --
Israel Binational Science Foundation under Research Grant Agreement
90-00483/2, by the German-Israeli Foundation for Scientific Research and
Development, by the Fund for Promotion of Research at the Technion, by the
N. S. E. R. C. of Canada and les Fonds F. C. A. R. du Qu\'ebec, and by the
United States Department of Energy under Contract No. DE FG02 90ER40560.

\newpage

\centerline{\bf APPENDIX: SU(3) REDUCED MATRIX ELEMENTS}
\bigskip
We relate the SU(3) reduced matrix elements as introduced by Zeppenfeld [8]
to the diagrammatic contributions described in Sec.~II.

We introduce a shorthand based on the decomposition (4)--(6) in Sec.~II.
There is a unique singlet amplitude; we denote it by \{1\}. The three octet
amplitudes arising from the ${\bf 3^*},~{\bf 6}$, and ${\bf 15^*}$
operators in (4)--(6) are denoted by \{8$_1$\}, \{8$_2$\}, and \{8$_3$\},
respectively. There is a unique amplitude \{27\}. The singlet and first
octet receive contributions involving both $V_{uq}V^*_{ub}$ and
$V_{tq}V^*_{tb}$ (we can eliminate $V_{cq}V^*_{cb}$ using unitarity), while
the second and third octets and the ${\bf 27}$-plet receive only
contributions proportional to $V_{uq}V^*_{ub}$. Here $q$ stands for $d$ or
$s$.

It is sufficient to discuss the case of $\Delta C = 0,~\Delta S = 0$
decays; a corresponding set of relations exists for the
strangeness-changing $\Delta C = 0$ amplitudes. Absorbing CKM factors into
the definitions of reduced amplitudes, we then have the following
relations:
$$
\{1\} = 2\st \left[ PA + \frac{2}{3}E + \frac{2}{3}P - \frac{1}{12}C
+ \frac{1}{4}T \right]~~~,
\eqno(A.1)
$$
$$
\{8_1\} = - \sqrt{\frac{5}{3}} \left[ P + \frac{3}{8}(T+A) - \frac{1}{8}
(C + E) \right]~~~,
\eqno(A.2)
$$
$$
\{8_2\} = \frac{\sqrt{5}}{4}(C + A - T - E)~~~,
\eqno(A.3)
$$
$$
\{8_3\} = - \frac{1}{8\sqrt{3}}(T + C) - \frac{5}{8\sqrt{3}}(A + E)~~~,
\eqno(A.4)
$$
$$
\{27\} = -\frac{T+C}{2\st}~~~.
\eqno(A.5)
$$
All amplitudes are linear combinations of these contributions. A
corresponding set of relations exists for the primed quantities, with
primed contributions related to unprimed ones by Eqs.~(\ref{ckmu}) and
(\ref{ckmt}).

The singlet amplitude is the only one which contains the penguin
annihilation ($PA$) contribution. It does not receive any contribution from
the annihilation graph, which contributes only to direct-channel octet
amplitudes. The penguin contribution $P$ appears only in the singlet and
first octet amplitudes.

If one neglects $E,~A$, and $PA$ one obtains the relations (\ref{rel}),
reducing the number of independent SU(3) amplitudes from five to three.
However, the same relations also follow from the less restrictive
assumptions $E + A = E + PA = 0$. As one sees from the discussion of Tables
1 and 2, these are the only two independent combinations which contain
exclusively contributions of graphs we wish to neglect.
\newpage

\centerline{\bf REFERENCES}
\begin{enumerate}

\item[{[1]}] D. London and R. Peccei, Phys.~Lett.~B {\bf 223}, 257 (1989);
M. Gronau, Phys.~Rev.~Lett.~{\bf 63}, 1451 (1989); B. Grinstein,
Phys.~Lett.~B {\bf 229}, 280 (1989); M. Gronau, Phys.~Lett.~B {\bf 300},
163 (1993).

\item[{[2]}] M. Gronau and D. London, Phys.~Rev.~Lett.~{\bf 65}, 3381
(1990).

\item[{[3]}] Yosef Nir and Helen R. Quinn, Phys.~Rev.~Lett.~{\bf 67},
541 (1991).

\item[{[4]}] Michael Gronau, Phys.~Lett.~B {\bf 265}, 389 (1991);
L. Lavoura, Mod.~Phys.~Lett.~A {\bf 17}, 1553 (1992).

\item[{[5]}] H. J. Lipkin, Y. Nir, H. R. Quinn and A. E. Snyder,
Phys.~Rev.~D {\bf 44}, 1454 (1991).

\item[{[6]}] J. Adler {\it et al.}, Phys.~Lett.~B {\bf 196}, 107 (1987);
S. Stone, in {\it Heavy Flavours}, edited by A. J. Buras and
M. Lindner (Singapore, World Scientific, 1992), p.~334;
J. C. Anjos {\it et al.}, Phys.~Rev.~D {\bf 48}, 56 (1993).

\item[{[7]}] D. Besson, invited talk at International Symposium on
Lepton and Photon Interactions at High Energies, Cornell University,
August, 1993 (unpublished).

\item[{[8]}] D. Zeppenfeld, Z. Phys.~C {\bf 8}, 77 (1981).

\item[{[9]}] M. Savage and M. Wise, Phys.~Rev.~D {\bf 39}, 3346 (1989);
{\it ibid.}~{\bf 40}, 3127(E) (1989).

\item[{[10]}] L. L. Chau {\it et al.}, Phys.~Rev.~D {\bf 43}, 2176
(1991).

\item[{[11]}] J. Silva and L. Wolfenstein, Phys.~Rev. D {\bf 49}, R1151 (1994).

\item[{[12]}] A. Deandrea {\it et al.}, Phys.~Lett.~B {\bf 320}, 170 (1994).

\item[{[13]}] Takemi Hayashi, M. Matsuda, and M. Tanimoto, Phys.~Lett.~B
{\bf 323}, 78 (1994).

\item[{[14]}] F. J. Gilman and R. Kauffman, Phys.~Rev.~D {\bf 36}, 2761
(1987); {\it ibid.} {\bf 37}, 3348(E) (1988).

\item[{[15]}] M. Battle {\it et al.} (CLEO Collaboration),
Phys.~Rev.~Lett.~{\bf 71}, 3922 (1993).

\item[{[16]}] T. Inami and C. S. Lim, Prog.~Theor.~Phys.~{\bf 65}, 297,
1772(E) (1981); G. Eilam and N. G. Deshpande, Phys.~Rev.~D {\bf 26}, 2463
(1982).

\item[{[17]}] See, e.g., Y. Nir and H. Quinn, Ann.~Rev.~Nucl.~Part.~Sci.~
{\bf 42}, 211 (1992).

\item[{[18]}] M. Bander, D. Silverman, and A. Soni, Phys.~Rev.~Lett.~{\bf 43},
242 (1979); G. Eilam, M. Gronau and J. L. Rosner, Phys.~Rev.~D {\bf 39}, 819
(1989); J.-M.~Gerard and W.-S.~Hou, Phys. Rev. D {\bf 43}, 2909 (1991).

\item[{[19]}] H. Yamamoto, Harvard Univ.~report HUTP-94/A006, 1994
(unpublished).

\item[{[20]}] H. Albrecht {\it et al.} (ARGUS Collaboration), Phys.~Lett.~
{\bf 158B}, 525 (1985); C. Bebek {\it et al.} (CLEO Collaboration),
Phys.~Rev.~Lett.~{\bf 56}, 1893 (1986).

\item[{[21]}] M. Gronau, J. L. Rosner, and D. London, Technion preprint
TECHNION-PH-94-7, March, 1994, submitted to Phys.~Rev.~Letters.

\item[{[22]}] O.F. Hern\'andez, D. London, M. Gronau, and J. L. Rosner,
Universit\'e de Montr\'eal preprint UdeM-LPN-TH-94-195, April, 1994,
submitted to Physics Letters.

\item[{[23]}] M. Gronau and D. Wyler, Phys.~Lett.~B {\bf 265}, 172 (1991).

\item[{[24]}] H. J. Lipkin, Phys.~Lett.~B {\bf 283}, 421 (1992).

\item[{[25]}] H. J. Lipkin, Phys.~Lett.~B {\bf 254}, 247 (1991).

\item[{[26]}] R. L. Kingsley, S. B. Treiman, F. Wilczek, and A. Zee,
Phys.~Rev.~D {\bf 11}, 1919 (1975); M. B. Einhorn and C. Quigg, {\it ibid.}
{\bf 12}, 2015 (1975); L.-L. Wang and F. Wilczek, Phys.~Rev.~Lett.~{\bf 43},
816 (1979); C. Quigg, Z. Phys.~C {\bf 4}, 55 (1979).

\item[{[27]}] See also J. Alexander {\it et al.} (CLEO Collaboration),
Phys.~Rev.~Lett.~{\bf 65}, 1184 (1990), where this ratio is quoted as
$2.35 \pm 0.37 \pm 0.28$;  M.~Selen {\it et al.} (CLEO Collaboration),
Phys.~ Rev.~Lett.~{\bf 71}, 1973 (1993), containing a more precise
measurement of $D^0 \to \pi^+ \pi^-$.

\item[{[28]}] V. M. Belyaev, A. Khodjamirian and R. R\"uckl, CERN
preprint CERN-TH-6880-93, May, 1993.

\end{enumerate}

\newpage

\centerline{\bf FIGURE CAPTIONS}
\bigskip

\noindent
FIG.~1. Diagrams describing decays of $B$ mesons to pairs of light
pseudoscalar mesons. Here $\bar q = \bar d$ for unprimed amplitudes and
$\bar s$ for primed amplitudes. (a) ``Tree'' (color-favored) amplitude $T$
or $T'$; (b) ``Color-suppressed'' amplitude $C$ or $C'$; (c) ``Penguin''
amplitude $P$ or $P'$ (we do not show intermediate quarks and gluons); (d)
``Exchange'' amplitude $E$ or $E'$; (e) ``Annihilation'' amplitude $A$ or
$A'$; (f) ``Penguin annihilation'' amplitude $PA$ or $PA'$.
\bigskip

\noindent
FIG.~2. Isospin triangles for $B \to \pi \pi$ (upper) and $\bar B \to \pi
\pi$ (lower). The lower triangle can also be flipped about the horizontal
axis. Here $A^{+-} \equiv A(B^0 \to \pi^+ \pi^-)$, $A^{00} \equiv A(B^0 \to
\pi^0 \pi^0)$, $A^{+0} \equiv A(B^+ \to \pi^+ \pi^0)$, $\tilde{A}^{+-} =
\tilde{A}(\Bbar^0 \to \pi^+ \pi^-)$, $\tilde{A}^{00} \equiv
\tilde{A}(\Bbar^0 \to \pi^0 \pi^0)$, $\tilde{A}^{-0} \equiv \tilde{A}(B^-
\to \pi^- \pi^0)$. The isospin amplitudes $a_2$, $A_0$, and $\tilde{A}_0$
are defined in the text.
\bigskip

\noindent
FIG.~3. SU(3) triangles involving decays $B^{\pm} \to \pi^{\pm} \pi^0$ and
$B^+ \to \pi K$ (upper) or $B^- \to \pi K$ (lower). The lower triangle can
also be flipped about the horizontal axis. Here $A^{0+} \equiv A(B^+ \to
\pi^0 K^+)$, $A^{+0} \equiv A(B^+ \to \pi^+ K^0)$, $\tilde{A}^{0-} \equiv
\tilde{A}(B^- \to \pi^0K^-)$, $\tilde{A}^{-0} \equiv \tilde{A}(B^- \to
\pi^- \bar K^0)$. The isospin amplitudes $A_{3/2}$, $C_{1/2}$, and
$\tilde{C}_{1/2}$ are defined in the text.
\bigskip

\noindent
FIG.~4. SU(3) triangles involving decays of charged $B$'s which may be used
to measure the angle $\gamma$.  Here $A^{0+} \equiv A(B^+ \to \pi^0 K^+)$,
$A^{+0} \equiv A(B^+ \to \pi^+ K^0)$, $A^{0-} \equiv A(B^- \to \pi^0K^-)$,
$A^{-0} \equiv A(B^- \to \pi^- \bar K^0)$, $A_{\pi \pi}^{+0} \equiv A(B^+
\to \pi^+ \pi^0) = -(C+T)/\sqrt{2} = 3 a_2 e^{i \gamma} e^{i
\delta_2}/\sqrt{2}$, $A_{\pi \pi}^{-0} \equiv A(B^- \to \pi^- \pi^0) = 3
a_2 e^{-i \gamma} e^{i \delta_2}/\sqrt{2}$.
\bigskip

\noindent
FIG.~5. Triangle relations from which weak phases and strong phase shift
differences can be obtained in the limit of neglect of certain diagrams.
The black dot corresponds to the solution for the vertex of the triangle in
the relation (\ref{cttri}). (a) Relation based on (\ref{tb}) (upper
triangle) and (\ref{tc}) (lower triangle). (b)  Relation based on
(\ref{taa}) (lower triangle with small circle about its vertex) and
(\ref{tb}) (upper triangle with large circle about its vertex).
\bigskip

\noindent
FIG.~6. Isospin triangle for $B \to \pi \pi$ decays with a small circle of
uncertainty associated with unknown final-state phase of the penguin
amplitude $P$. Here $C + T = -\s A(B^+ \to \pi^+ \pi^0)$, $T + P = -A(B^0
\to \pi^+ \pi^-)$, and $C - P = - \s A(B^0 \to \pi^0 \pi^0)$.

\end{document}